\documentclass[review]{elsarticle}

\usepackage{lineno,hyperref}
\modulolinenumbers[5]

\journal{Astroparticle Physics}

\begin{document}

\begin{frontmatter}

\title{Characterization of the background for a neutrino search with the HAWC observatory}

\author[1]{A. Albert}
\author[2]{R. Alfaro}
\author[3]{C.~Alvarez}
\author[2]{J.R. Angeles Camacho}
\author[4]{J.C.~Arteaga-Vel\'azquez}
\author[5]{K.P.~Arunbabu}
\author[2]{E.~Belmont-Moreno}
\author[3]{K.S.~Caballero-Mora}
\author[7]{T.~Capistr\'an}
\author[8]{A.~Carrami\~nana}
\author[9]{S.~Casanova}
\author[4]{U.~Cotti}
\author[10]{J.~Cotzomi}
\author[8]{S.~Couti\~no de Le\'on}
\author[11,12]{E.~De la Fuente}
\author[8]{R.~Diaz Hernandez}
\author[13]{M. A. DuVernois}
\author[1]{M.~Durocher}
\author[2]{C.~Espinoza}
\author[14]{K.L.~Fan}
\author[7]{N.~Fraija}
\author[2]{D.~Garcia}
\author[15]{J.A.~Garc\'ia-Gonz\'alez}
\author[7]{F.~Garfias}
\author[7]{M.M.~Gonz\'alez}
\author[14]{J.A.~Goodman}
\author[16]{D.~Huang}
\author[3]{F.~Hueyotl-Zahuantitla}
\author[16]{P.~H\"untemeyer}
\author[7]{A.~Iriarte}
\author[17,18,19]{A.~Jardin-Blicq}
\author[20]{D. Kieda}
\author[5]{A.~Lara}
\author[7]{W.H.~Lee}
\author[2]{H. Le\'on Vargas \corref{mycorrespondingauthor}}
\cortext[mycorrespondingauthor]{Corresponding author}
\ead{hleonvar@fisica.unam.mx}
\author[7]{A.L.~Longinotti}
\author[21]{G.~Luis-Raya}
\author[1]{K.~Malone}
\author[22]{J.~Mart\'inez-Castro}
\author[23]{J.A.~Matthews}
\author[24]{P.~Miranda-Romagnoli}
\author[4]{J.A.~Morales-Soto}
\author[10]{E.~Moreno}
\author[9]{A.~Nayerhoda}
\author[25]{L.~Nellen}
\author[24]{R.~Noriega-Papaqui}
\author[26]{N.~Omodei}
\author[27]{A.~Peisker}
\author[21]{E.G.~P\'erez-P\'erez}
\author[28]{C.D.~Rho}
\author[8]{D.~Rosa-Gonz\'alez}
\author[2]{A.~Sandoval}
\author[2]{J.~Serna-Franco}
\author[20]{R.W.~Springer}
\author[27]{K.~Tollefson}
\author[8]{I.~Torres}
\author[11,29]{R.~Torres-Escobedo}
\author[8]{F.~Ure\~na-Mena}
\author[10]{L.~Villase\~nor}
\author[29]{H.~Zhou}
\author[4]{C.~de Le\'on}

\address[1]{Physics Division, Los Alamos National Laboratory, Los Alamos, NM, USA}
\address[2]{Instituto de F\'isica, Universidad Nacional Aut\'onoma de M\'exico, Ciudad de M\'exico, M\'exico}
\address[3]{Universidad Aut\'onoma de Chiapas, Tuxtla Guti\'errez, Chiapas, M\'exico}
\address[4]{Universidad Michoacana de San Nicol\'as de Hidalgo, Morelia, Mexico }
\address[5]{Instituto de Geof\'{i}sica, Universidad Nacional Aut\'onoma de M\'exico, Ciudad de Mexico, Mexico }
\address[7]{Instituto de Astronom\'{i}a, Universidad Nacional Aut\'onoma de M\'exico, Ciudad de Mexico, Mexico }
\address[8]{Instituto Nacional de Astrof\'{i}sica, \'Optica y Electr\'onica, Puebla, Mexico }
\address[9]{Institute of Nuclear Physics Polish Academy of Sciences, PL-31342 IFJ-PAN, Krakow, Poland }
\address[10]{Facultad de Ciencias F\'{i}sico Matem\'aticas, Benem\'erita Universidad Aut\'onoma de Puebla, Puebla, Mexico }
\address[11]{Departamento de F\'{i}sica, Centro Universitario de Ciencias Exactas e Ingenierias, Universidad de Guadalajara, Guadalajara, Mexico }
\address[12]{Institute for Cosmic Ray Research, University of Tokyo, Kashiwa, Kashiwanoha, Japan}
\address[13]{Department of Physics and Wisconsin IceCube Particle Astrophysics Center, University of Wisconsin-Madison, Madison, WI, USA}
\address[14]{Department of Physics, University of Maryland, College Park, MD, USA }
\address[15]{Tecnologico de Monterrey, Escuela de Ingenier\'{i}a y Ciencias, Ave. Eugenio Garza Sada 2501, Monterrey, N.L., Mexico, 64849}
\address[16]{Department of Physics, Michigan Technological University, Houghton, MI, USA }
\address[17]{Max-Planck Institute for Nuclear Physics, 69117 Heidelberg, Germany}
\address[18]{Department of Physics, Faculty of Science, Chulalongkorn University, 254 Phayathai Road,Pathumwan, Bangkok 10330, Thailand}
\address[19]{National Astronomical Research Institute of Thailand (Public Organization), Don Kaeo, MaeRim, Chiang Mai 50180, Thailand}
\address[20]{Department of Physics and Astronomy, University of Utah, Salt Lake City, UT, USA }
\address[21]{Universidad Politecnica de Pachuca, Pachuca, Hgo, Mexico }
\address[22]{Centro de Investigaci\'on en Computaci\'on, Instituto Polit\'ecnico Nacional, M\'exico City, M\'exico.}
\address[23]{Dept of Physics and Astronomy, University of New Mexico, Albuquerque, NM, USA }
\address[24]{Universidad Aut\'onoma del Estado de Hidalgo, Pachuca, Mexico }
\address[25]{Instituto de Ciencias Nucleares, Universidad Nacional Aut\'onoma de Mexico, Ciudad de Mexico, Mexico }
\address[26]{Department of Physics, Stanford University: Stanford, CA 94305–4060, USA}
\address[27]{Department of Physics and Astronomy, Michigan State University, East Lansing, MI, USA }
\address[28]{University of Seoul, Seoul, Rep. of Korea}
\address[29]{Tsung-Dao Lee Institute \& School of Physics and Astronomy, Shanghai Jiao Tong University, Shanghai, People’s Republic of China}

%
%
%

\begin{abstract}
The close location of the HAWC observatory to the largest volcano in Mexico allows to perform a  search for neutrino-induced horizontal muon and tau charged leptons. The section of the volcano located at the horizon reaches values of slant depth larger than 8 km of rock, making it an excellent shield for the cosmic ray horizontal background. We report the search method and background suppression technique developed for the detection of Earth-skimming neutrinos with HAWC, as well as a model that describes the remaining background produced by scattered muons. We show that by increasing the detection energy threshold we could use HAWC to search for neutrino-induced charged leptons.
\end{abstract}

\begin{keyword}
Muons \sep Neutrinos \sep Earth-skimming 
\end{keyword}

\end{frontmatter}


\section{Introduction \label{sec::intro}}

Neutrinos are perhaps the most elusive particles of the Standard Model. They are probes that allow us to study weak interactions and the internal structure of nucleons and nuclei \cite{doi:10.1146/annurev-nucl-102115-044600}. At TeV energies, neutrinos allow us to test fundamental physics at energies that are not reachable at laboratories \cite{Ackermann:2019cxh} and those in the ultra high energy regime (UHE), starting at $\sim$100 TeV \cite{halzen}, point back to the most energetic particle accelerators in the Universe. In the GeV-TeV energy range neutrinos are predominantly of atmospheric origin.

Neutrino detection is a challenging task that requires special experimental conditions that range from building underground detectors, in order to avoid contamination from cosmic radiation, to requiring very large volumes of sensitive material such as ice or water to increase the probability of observing the visible signals from their weak interactions with matter. An example of such a detector is IceCube, with a 1 km$^{3}$ detection volume, that proved the existence of UHE astrophysical neutrinos \cite{2013IC}. 

Two decades ago, an alternative method called Earth-skimming was proposed to perform neutrino detection with above ground detectors \cite{Fargion:1999se,Feng:2001ue,Bertou_2002}. It consists of using mountains, or chords through the Earth, as a target to produce charged current neutrino-nucleon interactions. A number of early above-ground neutrino detectors are discussed in \cite{Spiering:2012xe}, but none have been built.

Several air shower detectors have been used to search for astrophysical neutrinos. For instance Ashra \cite{ASAOKA20137,Ashra:2019ICRC}, MAGIC \cite{Ahnen:2018ocv,Mallamaci:2019mmo} and the Pierre Auger observatory \cite{Aab:2019auo,2019ICRC:PA} have set upper limits on the neutrino flux at the UHE regime. Underground experiments have also searched for neutrinos associated to astrophysical sources, as for instance the detector of Crouch et al. \cite{Crouch:1978mg} Soudan 2 \cite{DeMuth_2004}, MACRO \cite{AMBROSIO20031}, Fr\'ejus \cite{RHODE1996217}, Kolar Gold Fields \cite{10.2307/78070}, LVD \cite{Alberini:1986ew} and Super-Kamiokande \cite{10.1093/ptep/ptab081}. HAWC presented preliminary results of a neutrino search in \cite{Vargas:2019kbb,Vargas:2021}. In this work we present a method to search for neutrino-induced charged leptons with a surface air shower array, providing evidence that it is feasible to use this type of detector to search for Earth-skimming neutrinos.

A limitation of the detection method discussed in this work is that only two of the three families of leptons can be studied in principle. The reason is that neutrino-induced electrons initiate electromagnetic showers shortly after production and are thus easily absorbed by the volcano. On the other hand, the large mean life of muons could allow them to escape the mountain and produce a measurable track within the HAWC array if the charged current interaction takes place close to the edge of the volcano to avoid absorption. Tau leptons have a mean life seven orders of magnitude shorter than that of muons and at energies below 10 PeV their survival probability to escape the volcano is dominated by their decay length \cite{Iyer_Dutta_2001}. If able to escape the volcano, a charged tau would produce a collimated air shower. Given the large dimensions of the HAWC WCDs the tau horizontal air shower might appear similar to a track in the detector array \cite{Vargas:2016hcp}. However, since tau-induced signals are more likely in the PeV energy range is safe to assume that all possible neutrino-induced candidates observed with HAWC are produced by muons.

\section{The HAWC observatory}
The High-Altitude Water Cherenkov (HAWC) observatory \cite{Abeysekara:2017mjj} is located at approximately 4100 m a.s.l. on the slopes of the Sierra Negra volcano in the state of Puebla, Mexico. The main array of HAWC consists of 300 Water Cherenkov Detectors (WCDs) distributed over a surface of 22,000 m$^2$. Each WCD is a cylindrical steel structure with a diameter of 7.3 m and 5 m tall. Inside of each of these structures is a plastic bladder that contains a water volume of approximately 200 000 litres. 
Each WCD is instrumented with four photomultiplier tubes (PMTs) fixed at its base.

The PMTs detect the Cherenkov light produced by charged particles as they pass through the water volume. Their calibration is performed using a laser system that sends pulses of different intensities in order to characterize the PMT response and to correct for the dependence of the timing with the light intensity, achieving nanosecond precision  \cite{2013ICRC:Calib}. Each individual PMT has a characteristic time offset due for instance to differences in cable lengths. These individual PMT time delays are characterized to sub nanosecond precision by fitting hadronic air showers \cite{Abeysekara:2017mjj}. All PMT pulses are digitized with the Time over Threshold (ToT) method with a double threshold with amplitudes of 1/4 and 4 photoelectrons (PEs). The signals are time-stamped with an accuracy of 100 ps and readout by a computer farm that continuously time orders the hits of the full HAWC array and generates an event trigger if there are at least 28 PMT hits within a sliding time window of 150 ns. A hit is any PMT signal that crosses the 1/4 PE threshold. The events are stored including all hits 0.5 $\mu$s before and 1 $\mu$s after the trigger for offline analysis \cite{Abeysekara:2017kao}.

\begin{figure}
\includegraphics[width=8.6cm]{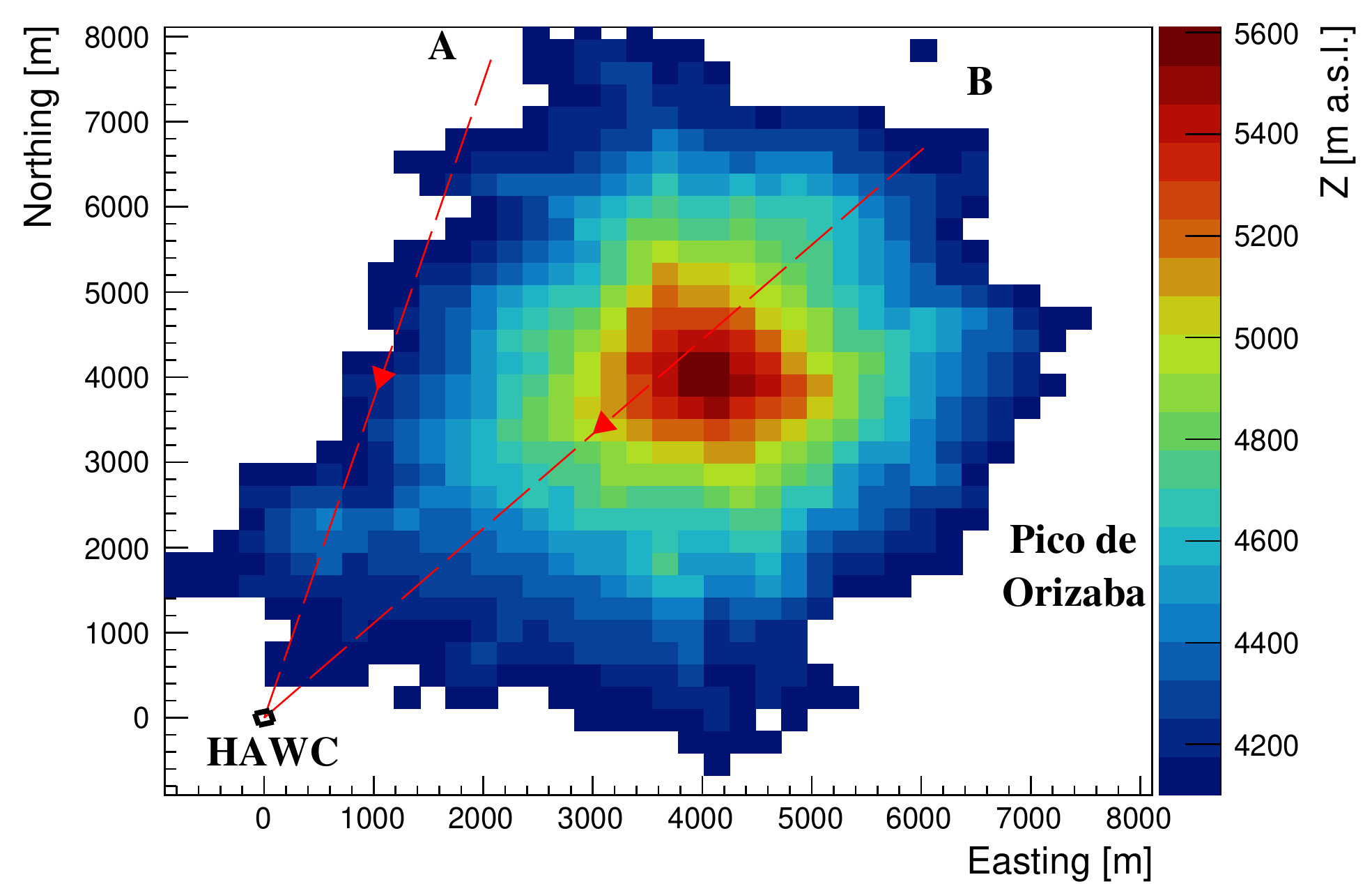}%
\caption{Elevation map of the Pico de Orizaba volcano. The observatory footprint is shown with a black square around the origin of the coordinate system. The color code indicates the altitude above sea level. Note that HAWC is located at $\approx$ 4100 m a.s.l. The arrows A and B show the approximate angular region (36$^\circ$ in azimuth) used in the analysis.\label{fig:topo}}
\end{figure}

The main objective of HAWC is to characterize gamma-ray sources in the energy range between a few hundred GeV up a few hundred TeV. In this work we use the Pico de Orizaba volcano as both target for neutrino-nucleon interactions and as an absorber for the horizontal atmospheric muon flux. Figure \ref{fig:topo} shows the topography of the volcano obtained with data from the Mexican National Institute of Statistics, Geography and Informatics (INEGI) \citep{inegi}. The location and approximate size of the HAWC observatory is shown by the area surrounded by black lines at the origin of the coordinate system. The arrows enclose the approximate angular region used in the analysis, from 282$^\circ$ to 318$^\circ$ in azimuth and up to 2$^\circ$ in the elevation direction as will be discussed in detail below. One can notice in Figure \ref{fig:topo} that the analysis region is not centered around the summit of the volcano. The reason for this is that the irregular distribution of WCDs in the HAWC array produces a strong dependence of the detector response with the azimuth angle. The analysis region was selected to optimize the detector performance as described below.

The method used for the neutrino search with HAWC is based on measuring the passage of the charged lepton (in the case of muons) produced in a charged current neutrino-nucleon interaction, or its highly collimated decay products (in the case of taus) \cite{Vargas:2016hcp}. This method is possible due to the modular design of the observatory that allows to use HAWC as a horizontal particle tracker. 

\begin{figure}
\includegraphics[width=8.6cm]{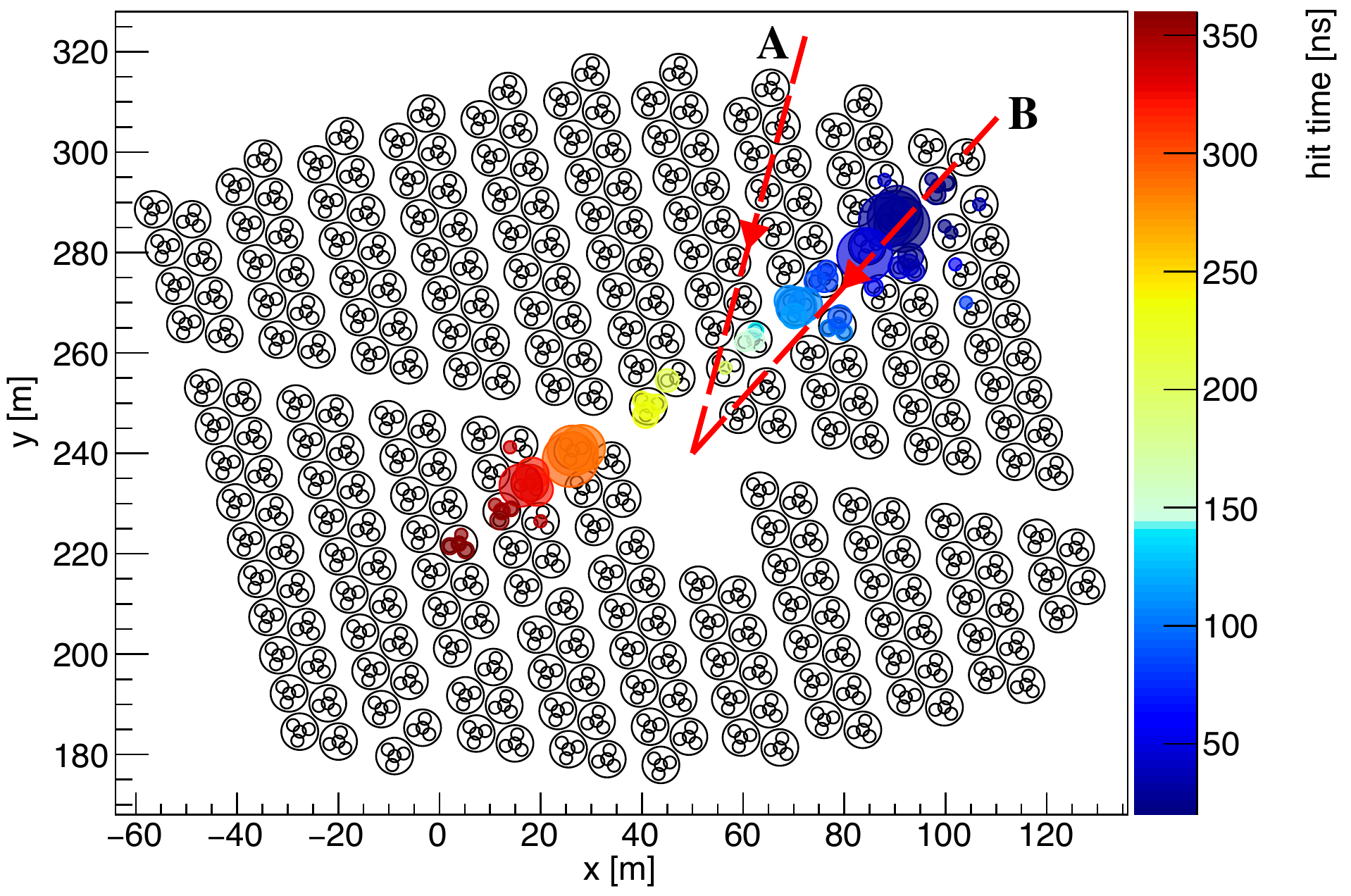}%
\caption{Simulation of a horizontal muon passing through HAWC. The color code indicates the PMT hit time while the size of the filled circles is proportional to the measured charge. The muon track points to the thickest region of the Pico de Orizaba volcano. The arrows show the approximate angular region used in the analysis and correspond to those shown in Figure \ref{fig:topo}. The $x$ and $y$ axis are oriented along the Easting and Northing directions respectively.\label{fig:simMu}}
\end{figure}

Figure \ref{fig:simMu} shows the simulated detector response to an almost horizontal 10 TeV muon. The simulations presented in this work were prepared using the GEANT4 \cite{Agostinelli:2002hh} based framework of HAWC. In Figure \ref{fig:simMu} each WCD is represented by a circle that encloses four smaller circles that depict the PMTs located inside. The time coordinate is shown with a color scale and the size of each filled circle is proportional to the magnitude of the measured charge. To ease the visualization this particular simulation does not include the noise primarily caused by vertical muons in data. However, one can notice WCDs with signals that are not contained in a straight line of WCDs. These additional hits are caused by secondary particles produced during the muon propagation, as predicted by the simulations. The time needed for the simulated muon to propagate through the detector with this trajectory is $\sim$ 400 ns. Due to the excellent time resolution of the instrument we can infer the propagation direction even between pairs of WCDs and verify that the particle propagates at light velocity.

The HAWC multiplicity trigger (28 PMT hits in a 150 ns time window \cite{Abeysekara:2017kao}) is general enough to capture the signals of interest for this study. This is because the detector array is being constantly hit by both small air showers and single vertical muons. This produces a baseline hit rate where the hits from the muon track are added to in order to produce a trigger. Below it is shown that the trigger efficiency for horizontal muons is low, but still capable of producing enough triggers to perform this study. However, it is reasonable to look for tracks in events where only a fraction of the total number of available PMTs are hit, in order to avoid the contamination of a horizontal muon track with secondary particles from a gamma-ray or a cosmic-ray induced air shower that might appear in the same time window. This analysis is restricted to events that contain at most 100 (out of a maximum of $\approx$ 1200) PMT hits in a 1.5 $\mu$s time window, i.e. a fraction of the whole dataset obtained with the air shower trigger.

The data selection was based on a proxy for the detector stability: the number of consecutive subruns associated to each run. The HAWC data are archived in runs that have a maximum duration of 24 hours. Each run is subdivided in $\sim$ 2 minute subruns, giving a maximum of 692 subruns per run. For this analysis we selected runs with at least 400 consecutive subruns. We studied data acquired during the months of June to October, 2017 and from January to May, 2018. This data selection requirement reduced the live time to approximately six months. 

The processing of the data begins with the standard HAWC reconstruction algorithms that use the charge and time of the PMT hits to determine the event energy and direction \cite{Abeysekara:2017mjj}. Our tracking algorithm, described in the next section, is then run over all the PMT hits in the events.

\section{Tracking algorithm}

For each triggered event, the PMT hits are time ordered. We use a lookup table to find if a given PMT hit is followed temporally by another signal in a different PMT in the spatial vicinity of the first signal, activated within a time window constrained by speed of light propagation. Based on full Monte Carlo (MC) simulations we optimized the width of the time window, allowing a 40 ns uncertainty due to the different paths light can take inside the WCDs to reach the PMTs. The search process continues until no further hits are found that follow a sequence of neighbouring WCDs. 

We use a charge threshold of 4 PEs in the PMT signal in order to minimize noise contamination. This reduces the single PMT rates by a factor of $\sim 4$.  A minimum of two PMT hits with charge above the four PE threshold are required in order to consider a WCD as active for track reconstruction. A track candidate is saved for further processing if it comprises at least nine PMT hits, setting the minimum length of a track to three WCDs. A more strict verification of speed of light propagation is performed to the track candidate, as well as a verification that the positions of the WCDs involved in the track can be fit using a linear function (for the muon energies that we detect the effect of the Lorentz force is negligible \cite{Butikofer2018}); if this step is successful the track properties are calculated. 

The angular properties of the track are calculated using the survey measurements of each WCD. The azimuth angle $\phi$ is calculated in the $x$-$y$ plane (see Figure \ref{fig:simMu}), by fitting a straight line that passes through the centres of the first and last WCD considered for the track reconstruction. We select only the signals whose reconstructed trajectory intersects at least 75\% of the WCDs that were identified as part of the track in the previous step. The intersection of the track with each WCD in the $x$-$y$ plane should be of at least 4 m. We follow the convention of measuring $\phi$ starting from East and in clockwise direction. 
The elevation angle $\theta$ is obtained by combining the height and $x$-$y$  positions of the WCDs that participate in the track. HAWC is located on a flat surface that limits the elevation angles that are possible to study. In this work we only use tracks with reconstructed elevation $< 2^{\circ}$. At present we are not able to distinguish if a muon is propagating upwards or downwards within the array. Both trajectories share the property of having a larger amount of Cherenkov light detected at the central WCDs of the track compared to that measured at the edges of the track. As it will be discussed below, our data is dominated by scattered muons with downward propagation. A solution to the upward-downward ambiguity will be needed in the future if high energy signals are detected and precise pointing is required to look for associations with possible sources.

\begin{figure}
\includegraphics[width=8.6cm]{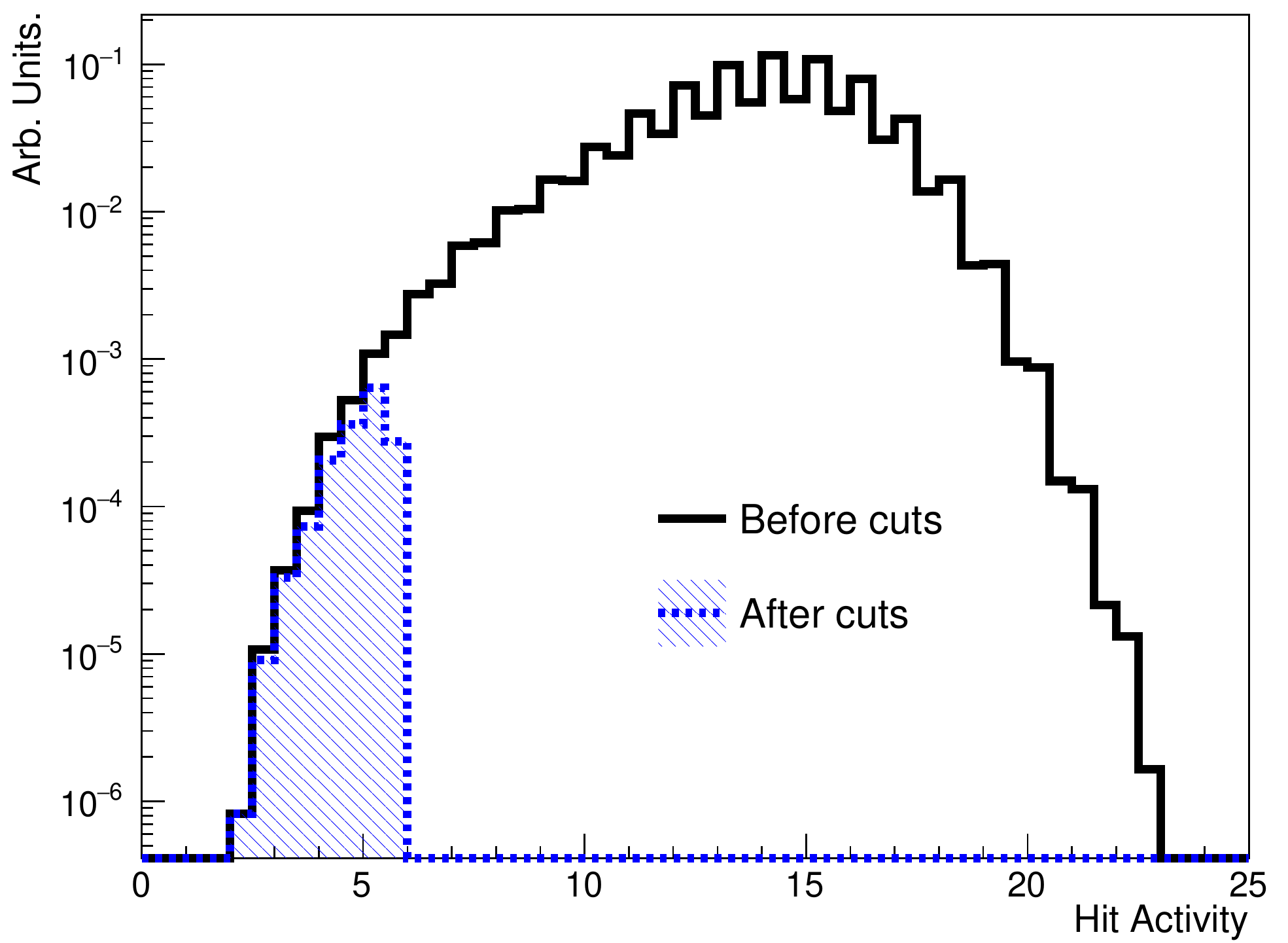}%
\caption{Distribution of values for the H$\mathrm{_A}$ variable. The distribution before applying the H$\mathrm{_A}$ and $\mathrm{M_{HA}}$ cut is normalized to unity. The shaded area shows the distribution of values for the track candidates after the cuts .\label{fig:hitactivity}}
\end{figure}

\begin{figure}
\includegraphics[width=8.6cm]{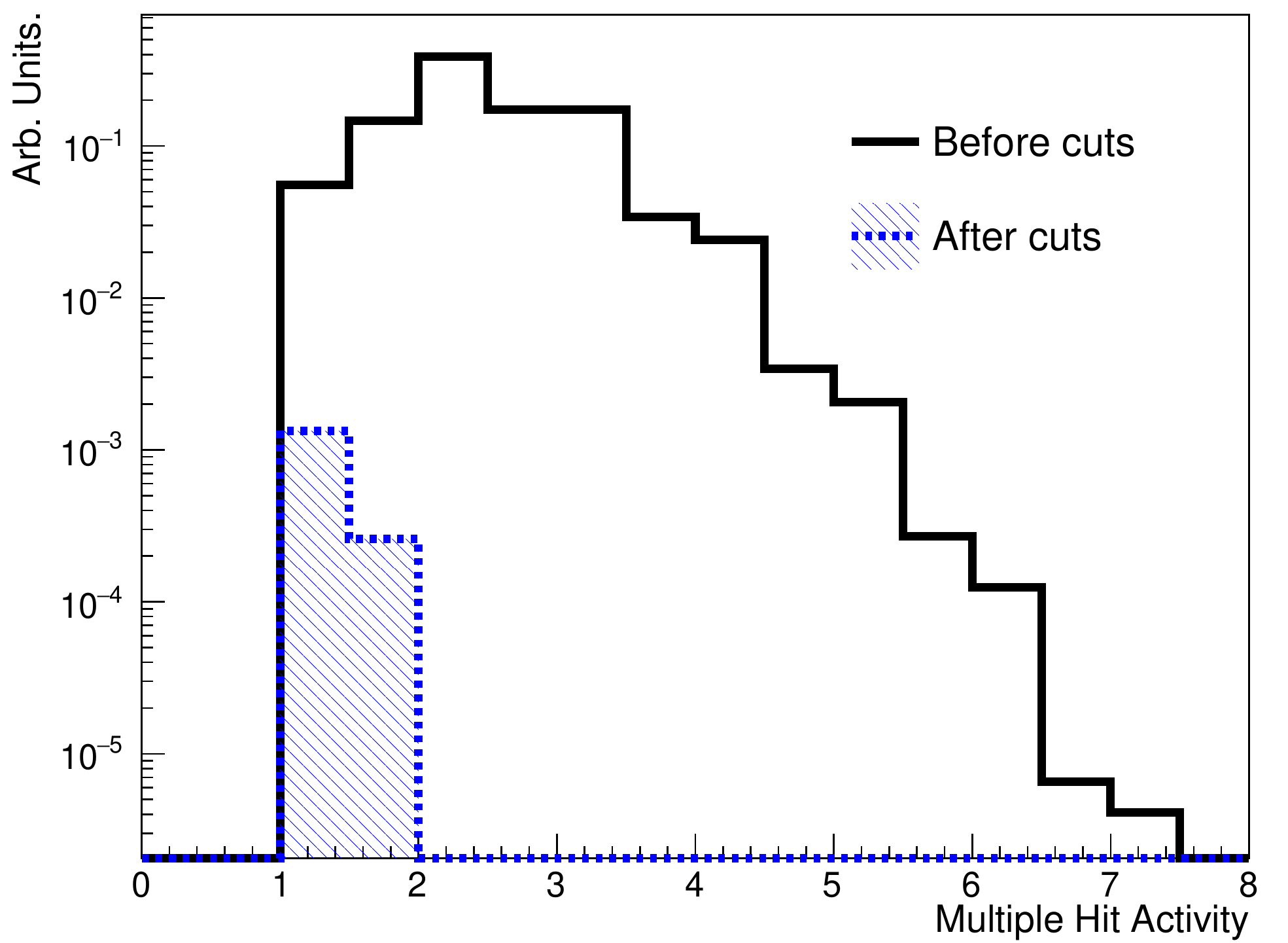}%
\caption{Distribution of values for the $\mathrm{M_{HA}}$ variable. The distribution before applying the $\mathrm{M_{HA}}$ and H$\mathrm{_A}$ cut is normalized to unity. The shaded area shows the distribution of values for the track candidates after the cuts .\label{fig:mhitactivity}}
\end{figure}

The final step in the processing of the track candidates consists in making very strict requirements on their isolation. This is done due to the presence of a huge background from very inclined air showers that are capable of satisfying the previously mentioned set of selection conditions. The values of the following additional cuts were determined by a detailed study of a large number of track candidates in data and with the use of full MC simulations. The simulated muons signals were embedded into a baseline of real data detector noise. We defined two variables that compare the number of WCDs (with PMT hits) associated to a track candidate with the total number of WCDs (with PMT hits) on each event:

\begin{itemize}
\item Hit Activity (H$\mathrm{_A}$): given by the ratio of the total number of WCDs with at least one PMT hit (without considering any charge threshold) in the 1.5 $\mu$s event window (N$_{\mathrm{WCDs}}$) and the track length within the detector volume, quantified by the number of WCDs that are associated to a track (N$\mathrm{_{WCDs}^{Track}}$), i.e. 
\begin{equation} \label{eq:AS}
\mathrm{H_A = N_{WCDs}/N_{WCDs}^{Track}},
\end{equation}
we cut on H$\mathrm{_A}$ $<$ 5.65 to remove air showers. By making a ratio we allow that triggered events with long tracks to contain more PMT hits that fall outside a straight line, as shown in Figure \ref{fig:simMu}, compared to shorter tracks. This cut reduces the number of track candidates by 99.83\%. Figure \ref{fig:hitactivity} shows the distribution of this variable in the data before and after applying these two selection cuts.

\item Multiple Hit Activity (M$\mathrm{_{HA}}$): given by the ratio of the total number of WCDs with at least two PMT hits with charge above 4 PEs ($\mathrm{N_{WCDs}^{M_{HA}}}$) in the event and the track length quantified by the number of WCDs that fulfil the previous requirement and are associated to the track, i.e.
\begin{equation} \label{eq:AH}
\mathrm{M_{HA} = N_{WCDs}^{M_{HA}}/N_{WCDs}^{Track}},
\end{equation}
the cut was set as M$\mathrm{_{HA}}$ $<$ 1.5. We expect that a horizontal muon will produce the largest fraction of PMT hits above the 4 PE threshold. However, we have to consider that there could be vertical muons in the same event that could also produce PMT hits that satisfy the charge threshold cut and thus activate additional WCDs. This is the reason why we allow up to $\approx$ 50\% more WCDs with multiple hit activity in the events. This cut reduces the number of track candidates by 96.13\%. Figure \ref{fig:mhitactivity} shows the distribution of this variable in the data before and after applying these two selection cuts.
\end{itemize}

The combination of both cuts reduces the track candidate sample by 99.93\%. These cuts are very strict and remove a significant fraction of horizontal muons according to the simulations, but we choose them because when using them we estimate a false positive rate smaller than 1\% .

\begin{figure}
\includegraphics[width=8.6cm]{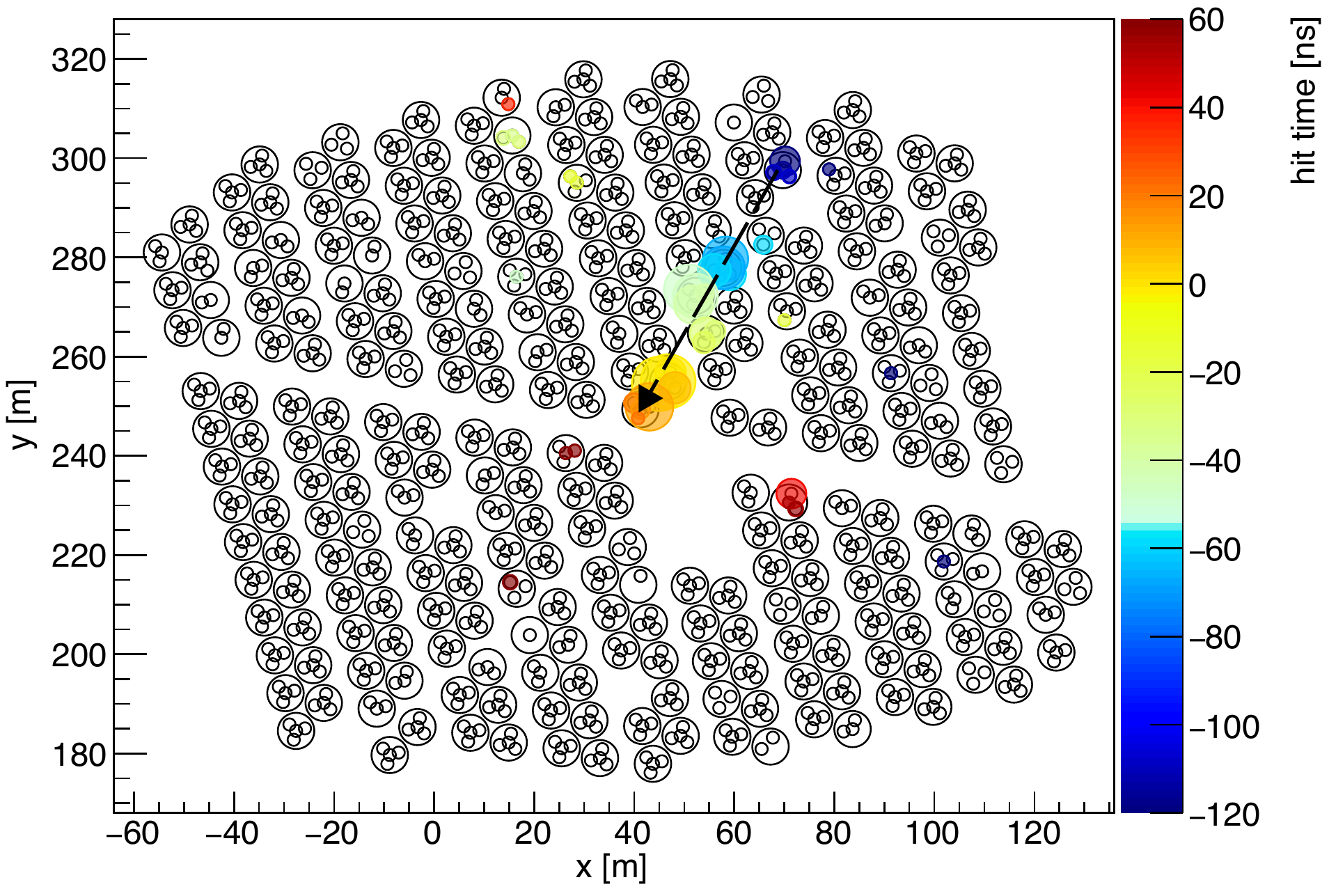}%
\caption{Near horizontal track found in the data whose direction points to the volcano. The color code indicates the time while the size of the filled circles is proportional to the charge.\label{fig:extrack}}
\end{figure}

Another possible source of contamination is artificial tracks created by combinatorial background. In order to assess its possible contribution we used a data-based method. For the whole data sample we randomized the PMT locations, keeping the corresponding charge and time information. Then we ran our tracking algorithm over this randomized data set. The algorithm was able to identify artificial tracks with an average number of 0.01\% of that from real track candidates, but over the whole six month live time the maximum length of these artificial tracks was of three WCDs. We decided to make an additional cut that required tracks to have a length equal or longer than four WCDs to completely remove the combinatorial background. Figure \ref{fig:extrack} shows an example of a track found in the data that points back to the Pico de Orizaba volcano.

The performance of the tracking algorithm was tested using simulated muons in the energy range [100 GeV, 5 TeV], embedded into noise obtained from untriggered data. The detector efficiency has a strong dependence on the track azimuth angle due to the irregular distribution of WCDs in the observatory. The regions between columns and rows of WCDs produce preferential directions where both the angular resolution degrades and the rate of artificial tracks due to near horizontal air showers is increased. There are, however, several regions of the acceptance pointing to the volcano where the angular resolution is reasonably good ($< 2^\circ$). The tracks presented on this work have a pointing accuracy $< 2^\circ$.

For this analysis we restrict our search of tracks to those that point back to the base of the Pico de Orizaba volcano, the region that provides the best shielding against atmospheric muons. We segmented the acceptance in rectangular bins of six degrees in azimuth and two degrees in elevation. By restricting our study to near horizontal tracks the measurements are less affected by reconstruction inefficiencies. We use for the analysis the six bins that have the best reconstruction efficiency (defined as the ratio between the number of reconstructed muon tracks and the number of MC generated muons) and angular resolution ([282$^{\circ}$, 318$^{\circ}$] in azimuth, delimited by the red lines in Figure \ref{fig:shadow1kmwe}). The angular resolution in the azimuth direction in this region is $<$ 2$^{\circ}$ with the exception of the interval [294$^{\circ}$, 300$^{\circ}$], where the angular resolution is $\approx$ 2.5$^{\circ}$.  This particular azimuth bin is however well contained by the full analysis region. The angular resolution in the elevation direction, for a reconstructed elevation $\theta_{\mathrm{Rec}}<2^{\circ}$ in the six analysis bins, is of 0.7$^{\circ}$. The angular resolution, as expected, degrades for increasing reconstructed elevation angles. For instance, for $4^{\circ}< \theta_{\mathrm{Rec}}<6^{\circ}$, the angular resolution is of 3.5$^{\circ}$.

After performing the analysis on the whole data sample we identified 122 muon track candidates that point back to the combined six analysis bins shown in Figure \ref{fig:shadow1kmwe}. As will be discussed below, this number of events can be explained as being produced by low energy muons that are scattered towards almost horizontal directions, as discussed in the next section. The results are presented for the combination of the six analysis bins, and also for the subregions I and II delimited by dashed lines in Figure \ref{fig:shadow1kmwe}. 

\section{Model of the background from scattered muons}

Once we are able to reject both the background from air showers and the combinatorial one, the largest remaining background for a neutrino search with a surface array is produced by muons that are scattered towards horizontal directions and seem to point back to the volcano \cite{Elbert:1991kd,GRIEDER2001305,BONECHI2020100038}. 

\begin{figure}
\includegraphics[width=8.6cm]{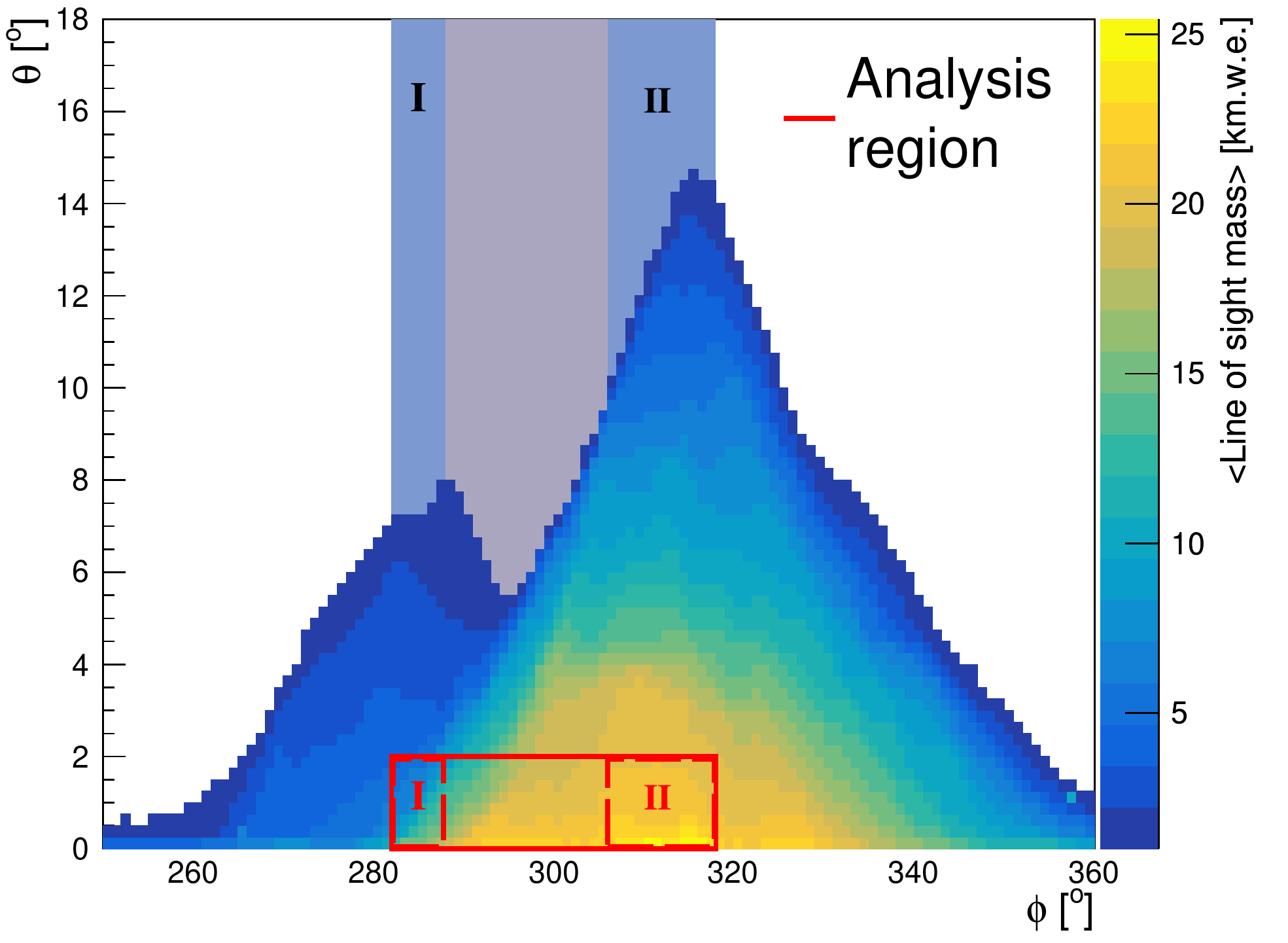}%
\caption{Pico de Orizaba profile as seen from the center of the HAWC main array. The shaded areas behind the mountain show the solid angles where muons propagate before being scattered towards almost horizontal directions into the three analysis regions, delimited by red lines (full region and regions I and II).  \label{fig:shadow1kmwe}}
\end{figure}

The volcano acts as a shield against the atmospheric muon flux. Therefore we need to study the effect that it produces on the muons that could be scattered. The shadow of the mountain is shown in Figure \ref{fig:shadow1kmwe}, that displays the values of the \textit{line of sight mass} (LOSM). The LOSM is the amount of matter, measured in km.w.e., that an atmospheric muon would need to travel through before reaching the HAWC observatory and being detected at a particular analysis bin. The color code shows values larger than 1 kilometre water equivalent (km.w.e.). The conversion from the rock overburden measured geometrically in km to km.w.e. was done considering the composition of the volcano. This large volcanic structure was built by different processes over a time span that comprises several thousands of years, producing eruptive products of different composition, which makes it difficult to determine a single and precise value for the density of the bulk of the Pico de Orizaba. However, for the purpose of this paper, we can use mean values of andesitic rocks to approximate the density of Pico de Orizaba with a value of 2.6 g/cm$^3$ \cite{Carrasco}. A study of the effects of the chemical composition on the attenuation of the muon flux can be found in \cite{se-9-1517-2018}, particularly useful in the context of muon tomography as in the recent results with the Khufu Pyramid \cite{cite-kufu}.

We model the background of muons that scatter towards almost horizontal directions using geometrical arguments. The shaded areas in Figure \ref{fig:shadow1kmwe} show the region above the volcano from where muons can travel unabsorbed towards HAWC and be scattered into the horizontal acceptance of the detector, faking their direction as coming from the analysis region. The darker shadings indicate the parts of the sky where muons can scatter into the analysis sub-regions I and II.

\begin{figure}
\includegraphics[width=8.6cm]{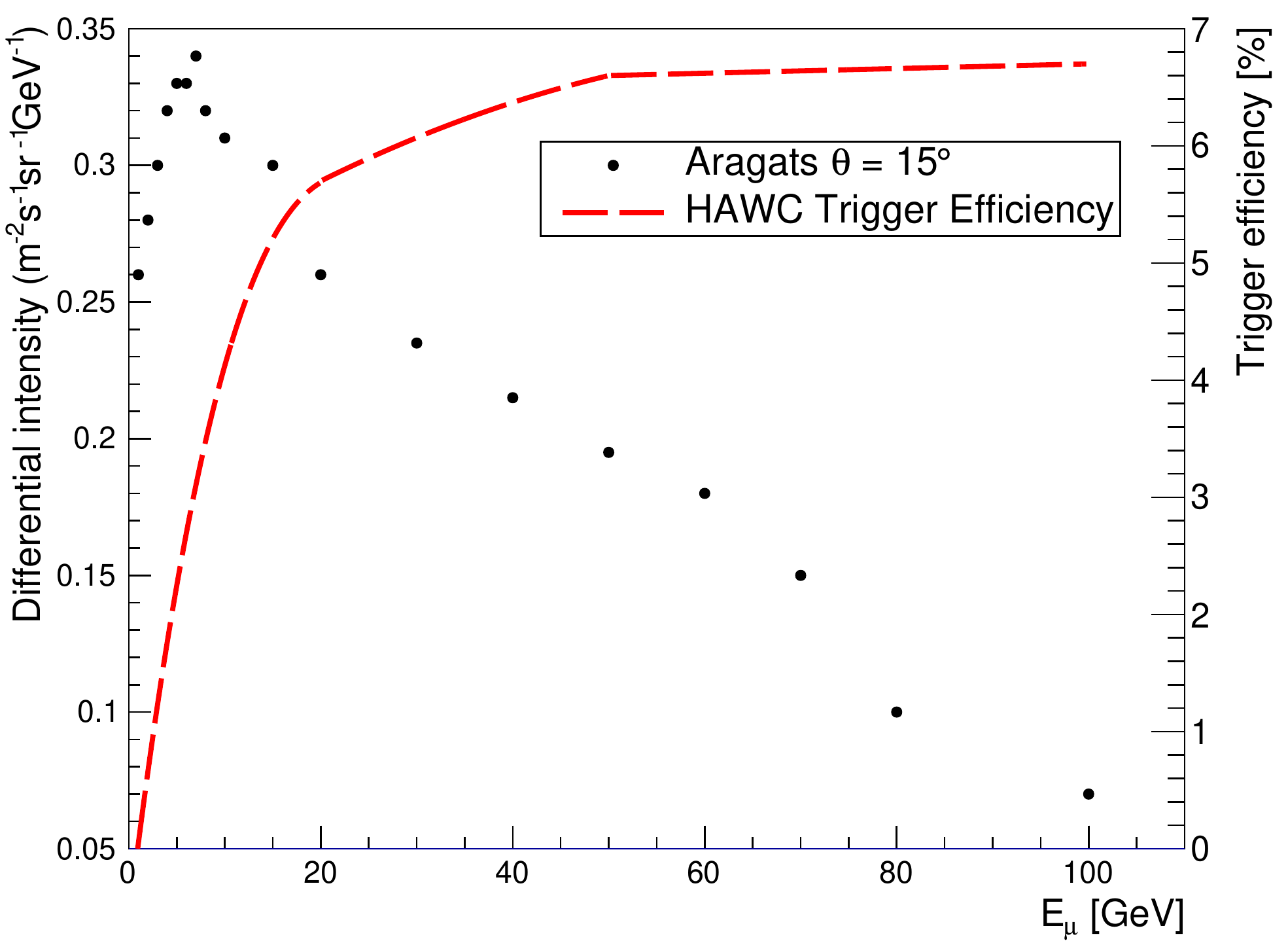}%
\caption{Differential intensity of muons from an elevation angle of $\theta$=15$^{\circ}$, shown as black dots, from the compilation from Aragats \cite{aragats}.  The HAWC trigger efficiency is shown as a red line with the right hand side scale.\label{fig:TE}}
\end{figure}

\begin{figure}
\includegraphics[width=8.6cm]{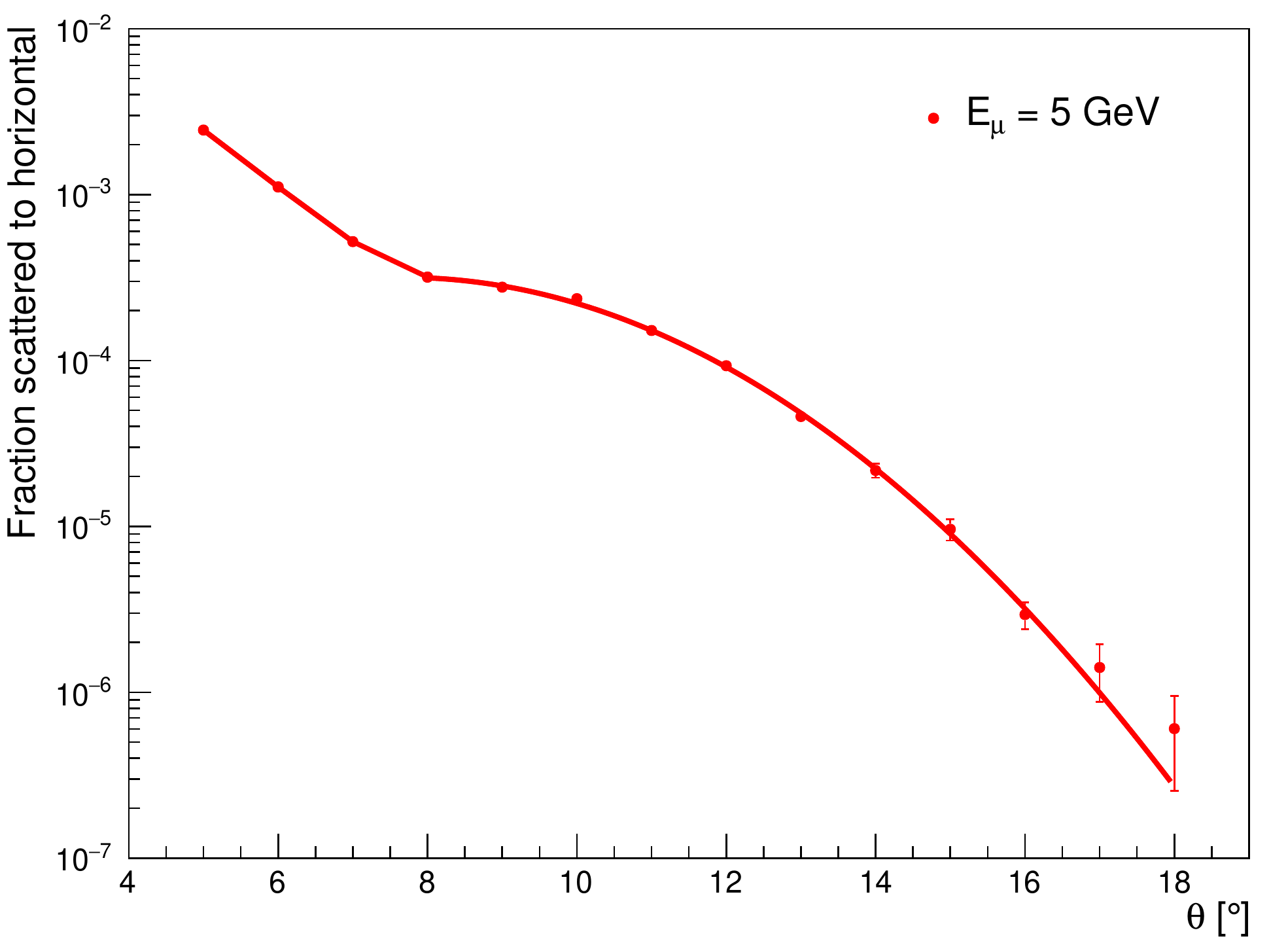}%
\caption{Fraction of simulated 5 GeV muons that are scattered towards almost horizontal directions as a function of the elevation angle. The solid line shows the fit used in the calculations. \label{fig:scattprob5GeV}}
\end{figure}

The atmospheric muon differential intensity as a function of the zenith angle close to the horizon was obtained by fitting the data available from \cite{aragats} at 3200 m a.s.l., which are the closest measurements to the HAWC altitude. In Figure \ref{fig:TE}, a sample of these data shows the differential intensity as a function of the muon energy at a fixed elevation angle of $\theta$=15$^{\circ}$. For muon energies that were not available in the data set from \cite{aragats} we used the results from the model presented in \cite{LIPARI1993195}.
Since the scattering probability is inversely proportional to the muon momentum \cite{BONECHI2020100038,Elbert:1991kd}, this background has a strong dependence on the muon energy. The scattering probability as a function of the muon energy and elevation angle was evaluated using the GEANT4 based simulation of the HAWC observatory. We calculated this probability by simulating an isotropic distribution of muons directed towards the HAWC observatory. This type of simulation is appropriate since the elevation dependence of the muon flux is taken into account by the fits to the Aragats \cite{aragats} measurements. We simulated muons with energies larger than 2 GeV since, according to our GEANT4 simulations, this was the lowest energy of muons that were able to be scattered and propagate at least through four WCDs. An example distribution of the scattering probability is shown in Figure \ref{fig:scattprob5GeV}, for 5 GeV muons. The solid line shows the fit to the simulation results used in the scattering calculation. The scatterings are almost completely dominated by muons that are deflected just enough to produce an almost horizontal trajectory, which based in our analysis strategy corresponds to trajectories that intersect at least four WCDs. 

The flux of muons that are scattered towards horizontal trajectories, for any 1$^{\circ}$ bin in azimuth, is calculated using Equation \ref{eq:flux1bin}
\begin{equation} \label{eq:flux1bin}
F^{\phi}_{\mathrm{Scatt}} (E) = \frac{\pi}{180} \int_{\theta_{i}}^{\theta_{f}} I (E,\theta) P_{\mathrm{Scatt}} (E,\theta) \sin\theta d\theta ,
\end{equation}
where $I(E,\theta)$ is the function that describes the muon differential intensity at energy $E$ close to the horizon and $P_{\mathrm{Scatt}}(E,\theta)$ is the function that describes the scattering probability at energy $E$ and elevation angle $\theta$ obtained from the fit to simulations as in Figure \ref{fig:scattprob5GeV}. The integration limits in the elevation angle ($\theta_{i}$ and $\theta_{f}$) are given by the geometrical constrains of the mountain profile for each particular azimuth bin, i.e. we evaluate the integral starting from the highest elevation covered by the mountain profile, $\theta_{i}$, and integrate up to $\theta_{f} =$ 18$^{\circ}$. We simplify the azimuth dependence of the scatterings by assuming that for any particular azimuth bin the probability that a muon can scatter in the azimuth direction and migrate to a neighbour bin cancels out due to the opposite effect happening, with the same probability, in the neighbour bin. We used our MC to evaluate the effect of this simplification and found that it produces a 3\% uncertainty in the scattering model. This is a small effect compared with the main uncertainty in the modelling caused by using different mountain profiles as will be discussed below.


We then add the contributions of the 1$^{\circ}$ bins in azimuth that go into each analysis region at a fixed energy $E$ with Equation \ref{eq:fluxAdd}
\begin{equation} \label{eq:fluxAdd}
F_{\mathrm{Scatt}} (E) = \sum_{\phi} F^{\phi}_{\mathrm{Scatt}} (E) ,
\end{equation}
and finally add the contribution from muons starting at 2 GeV (the approximate detection threshold with HAWC) up to 100 GeV. We divide the result by the corresponding solid angle covered by each analysis region, as shown in Equation \ref{eq:ScattInt}, to obtain the scattered intensity
\begin{equation} \label{eq:ScattInt}
\mathrm{I(i)}  = \sum_{E=2,3,...}^{E=100\; \mathrm{GeV}} F_{\mathrm{Scatt}} (E) /\Omega_{i} ,
\end{equation}
on the other hand the intensity from data is calculated using Equation \ref{eq:int}
\begin{equation} \label{eq:int}
\mathrm{I(i) = \frac{N^{Raw}_{i}}{\Delta T (A \Omega)_{i}} }  ,
\end{equation}
where $\mathrm{N^{Raw}_{i}}$ is the raw number of tracks of the i-th bin, $\mathrm{\Delta T}$ is the live time and $\mathrm{(A \Omega )_{i}}$ is the product of the area and solid angle. The solid angle $\mathrm{\Omega_{i}}$ corresponds to that covered by each analysis region. The effective area $\mathrm{A_{i}}$ for each analysis region was obtained with Equation \ref{eq:AE}
\begin{equation} \label{eq:AE}
\mathrm{A_{i} = \frac{N}{N_{Gen}} \times A_{Gen}} ,
\end{equation}
where $\mathrm{N}$ is the number of MC tracks that are detected by HAWC,  $\mathrm{N_{Gen}}$ is the number of generated tracks that point towards the observatory and whose trajectories intersect at least four WCDs, and $\mathrm{A_{Gen}}$ is the area over which the MC muons are generated. We use  MC simulations of muons embedded into real data noise. The detector noise comes from 1.5 $\mu$s samples of raw data that are verified so they do not produce air shower triggers. The noise is added to the GEANT4 simulation of individual muons before reconstructing the simulated events. The HAWC simulations also include a detailed implementation of the PMT-to-PMT relative variations in efficiency. Figure \ref{fig:TE} shows with a red line the trigger efficiency as a function of the muon energy. The detection efficiency increases with increasing muon energy due to the larger production of secondary particles. The additional PMT hits produced by the radiative energy loss of muons allows to increase the probability to reach the multiplicity trigger requirement. At 10 TeV the trigger efficiency is $>$ 18\%, compared to the $\approx$ 6.6\% at 60 GeV. 

The numerical values of the product $\mathrm{(A \Omega )_{i}}$ for the three analysis regions shown in Figure \ref{fig:shadow1kmwe} are reported in Table \ref{table:overburden}. These values correspond to the effective area of HAWC to detect muons with energies larger than 2 GeV with the analysis strategy described on this paper. The uncertainty in the effective area reported in Table \ref{table:overburden} is obtained by combining the statistical uncertainty in the MC and from a comparison of the muon intensity at 1.5 km.w.e. measured with HAWC from those from the Kolar Gold Mines experiment \cite{Kolar1977} and the compilation by Crouch \cite{Crouch1987}. Our measurement appears between that of those two previous experiments with a difference of 18\% with respect to each one. We assigned this difference as a systematic uncertainty of our effective area estimation. Table \ref{table:overburden} also shows the average LOSM values for the three selected analysis regions and the raw number of track candidates found on each analysis region. The uncertainty in the LOSM reported on Table \ref{table:overburden} corresponds to the maximum variation found in the distributions of values for each analysis region. The variation include the effect of changing the reference observation position throughout the HAWC platform. 

 \begin{table}
 \caption{Mean values of the \textit{line of sight mass} (LOSM) and effective area in the directions of interest shown in Figure \ref{fig:shadow1kmwe}. The column -tracks- indicates the raw number of signals detected in each region of the acceptance. \label{table:overburden}}
 \begin{tabular}{l c c c}
  Bin & $<$LOSM$>$ [km.w.e.] & A$\Omega$ [m$^2$sr] & Tracks   \\
  \hline
   Full & $17.70^{+3.35}_{-9.23}$ & 5.1$\times 10^{-2}$ $\pm$ 9.2$\times 10^{-3}$ & 122 \\
 \hline   
   I & $9.54^{+1.95}_{-1.92}$ & 4.1$\times 10^{-3}$ $\pm$ 7.5$\times 10^{-4}$  & 21 \\
 \hline   
   II & $20.97^{+0.09}_{-0.13}$ & 3.0$\times 10^{-2}$ $\pm$ 5.4$\times 10^{-3}$  & 26 \\
  \end{tabular}
 \end{table}

\begin{figure}
\includegraphics[width=8.6cm]{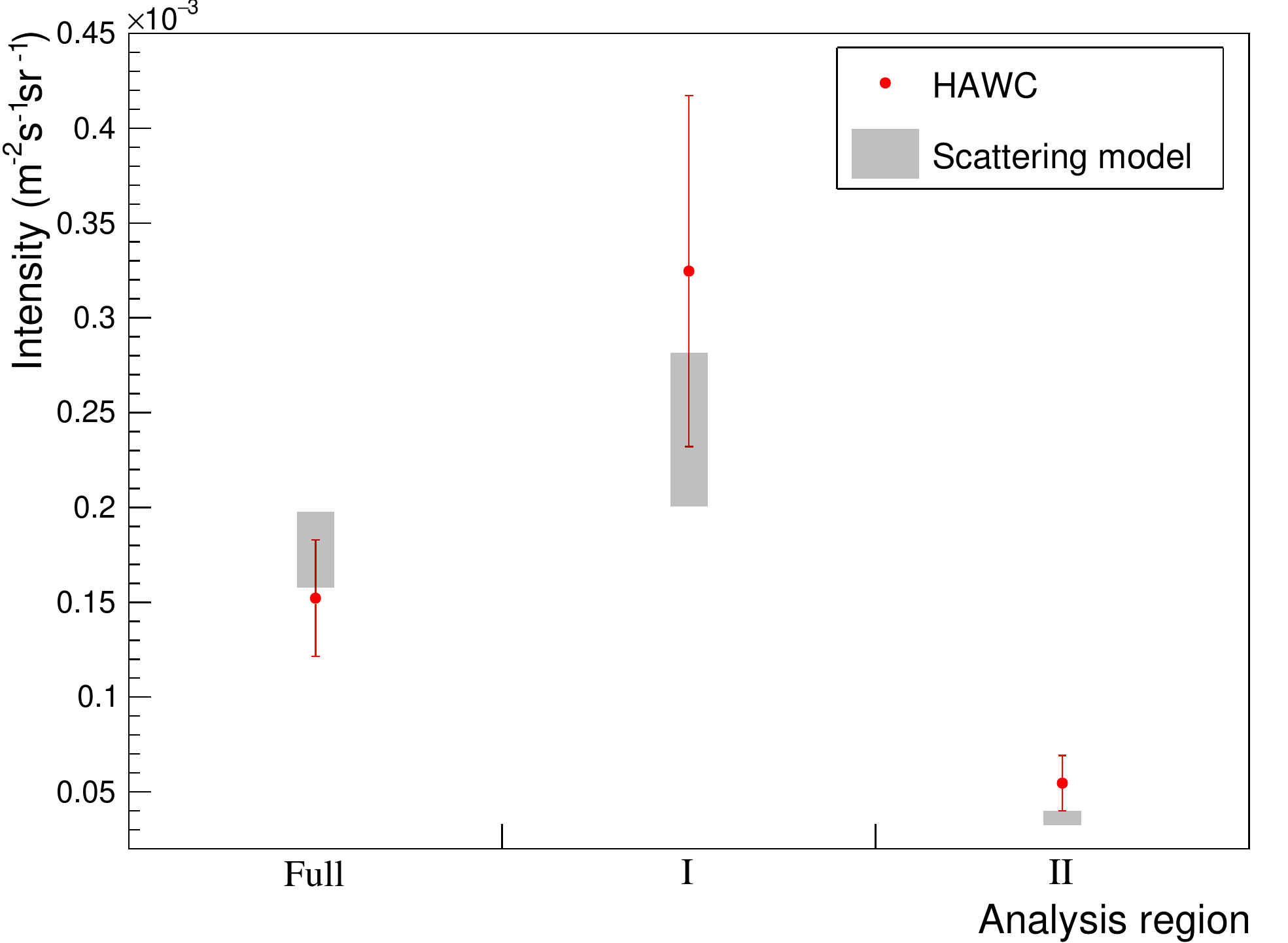}%
\caption{Muon intensity for the analysis bins that point back to the base of the Pico de Orizaba volcano. The shaded area shows the estimation of the scattered intensity in the corresponding regions.\label{fig:ScatteringLinear}}
\end{figure}

Figure \ref{fig:ScatteringLinear} shows the comparison of the measurement of HAWC with the result of the scattering model. There is fair agreement between the HAWC measured intensity and the scattering model, for instance the model is able to reproduce the approximate factor of six difference in intensity observed between the analysis region I and II. A more sophisticated estimation of the background for a neutrino search would include the propagation of muons through regions with low values ($<1$ km.w.e.) of the LOSM. However, such study is beyond the scope of this paper and we consider our simple modelling sufficient for our purposes given that our uncertainty in the intensity measurement currently ranges from 20\% to 30\%. As an uncertainty of our model we show in Figure \ref{fig:ScatteringLinear} how the expected scattered intensity changes by increasing the threshold value of the LOSM used to define the mountain profile from 1 to 2 km.w.e., the allowed range is shown with the shaded area. The uncertainty in the scattering model range from 10\% to 17\%.

\begin{figure}
\includegraphics[width=8.6cm]{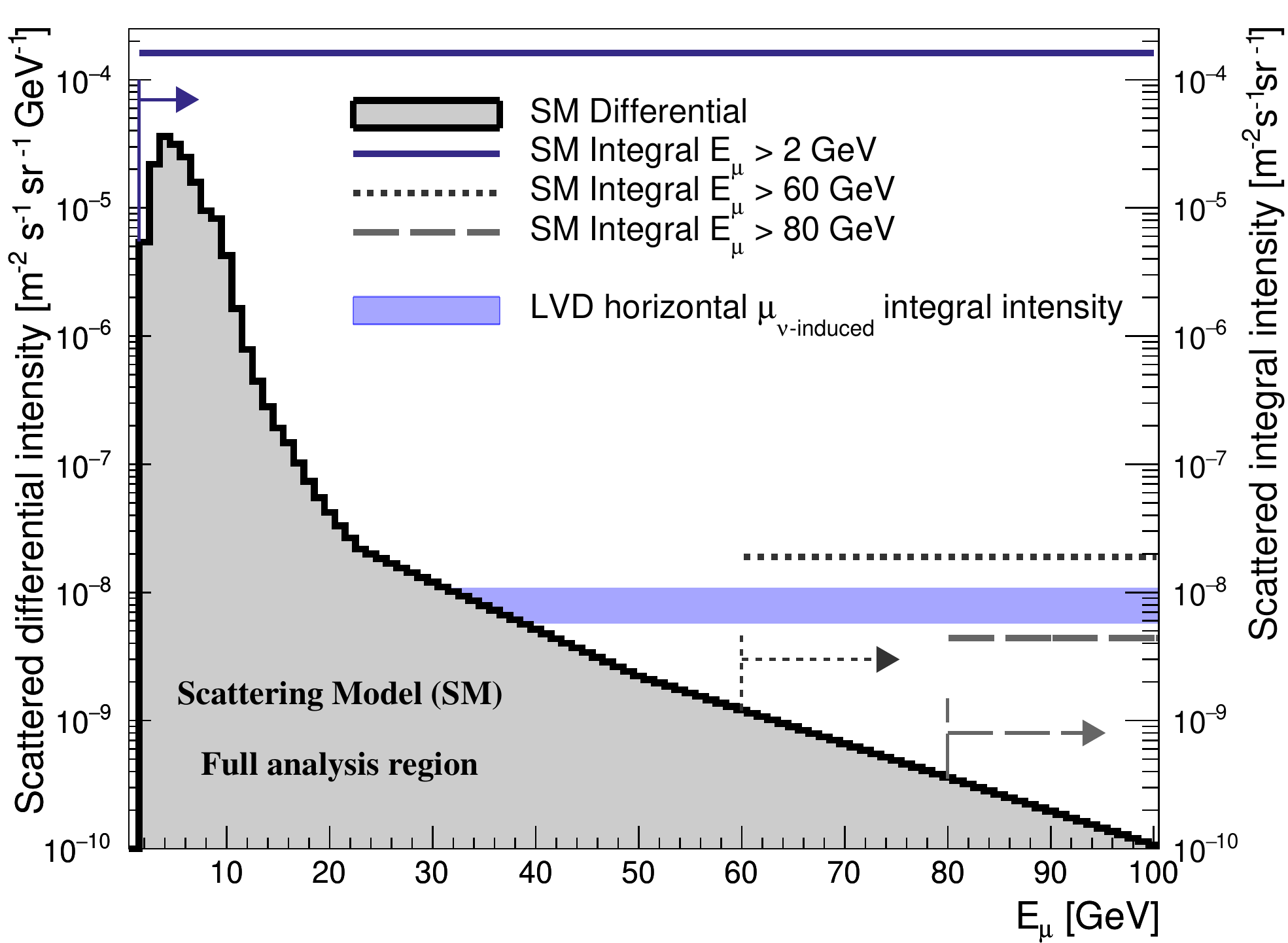}%
\caption{Simulated energy distribution of the scattered muons that point back to the full analysis region at the base of the Pico de Orizaba volcano. The horizontal lines show the integrated intensity for different muon energy thresholds according to our model. The arrows indicate the direction of the integration over the differential intensity distribution. The integral intensity of horizontal neutrino-induced muons measured by LVD \cite{AGLIETTA1995311} is shown as reference with a blue area. \label{fig:intensidadscatt}}
\end{figure}

Given the fair agreement between our scattering model and the measurement,  we can use the model to define a strategy to search for neutrino-induced muons. Figure \ref{fig:intensidadscatt} shows the predicted differential intensity of scattered muons as a function of the muon energy for the full analysis region. The solid horizontal line at the top of the figure shows the predicted integral intensity above the HAWC detection threshold which is consistent with our measurement from the full analysis region as shown in Figure \ref{fig:ScatteringLinear}. The dotted and dashed horizontal lines show the predicted integral intensity obtained when increasing the detection energy threshold to 60 GeV and to 80 GeV in our model. The blue shaded area shows, as a reference, the integral value of the neutrino-induced horizontal muon intensity measured by the LVD collaboration \cite{AGLIETTA1995311}. This analysis shows that by constraining the muon energy the scattering background can be reduced. For high energy muons ($>$ 100 GeV) the background is below the intensity of neutrino-induced muons in rock as measured by the LVD experiment, opening the possibility of their detection by the HAWC observatory using the Pico de Orizaba volcano as a neutrino target and atmospheric muon filter.

The energy estimation of near horizontal muons with HAWC is not a trivial task because we are only able to sample the energy deposited by a muon. We are exploring the possibility of using the independent charge measurement from different WCDs. The charge deposited in each WCD by the same muon shows fluctuations that depend on two factors: the proximity and incidence angle of the track with respect to the PMTs and fluctuations due to radiative energy losses \cite{Zyla:2020zbs,RADEL201253}. We plan to use the fact that high energy muons suffer stochastic energy losses that cause variations in the deposited charge in different WCDs along the muon trajectory. The dynamic range of the PMTs in HAWC goes from a fraction of a PE up to thousands of PEs, and an average atmospheric muon produces $\approx$ 30 PEs \cite{smith2015hawc}, leaving ample room to characterize large energy losses. Dedicated neutrino observatories such as IceCube have made use of catastrophic energy losses to identify very high energy muons \cite{AARTSEN20161}. In \cite{Vargas:2019kbb} we presented preliminary results of muon signals that produce very large deposits of charge in the detector. At this moment we are not yet able to provide reliable estimations of the muon energies and will leave such results as the subject of future publications. 

\begin{figure}
\includegraphics[width=8.6cm]{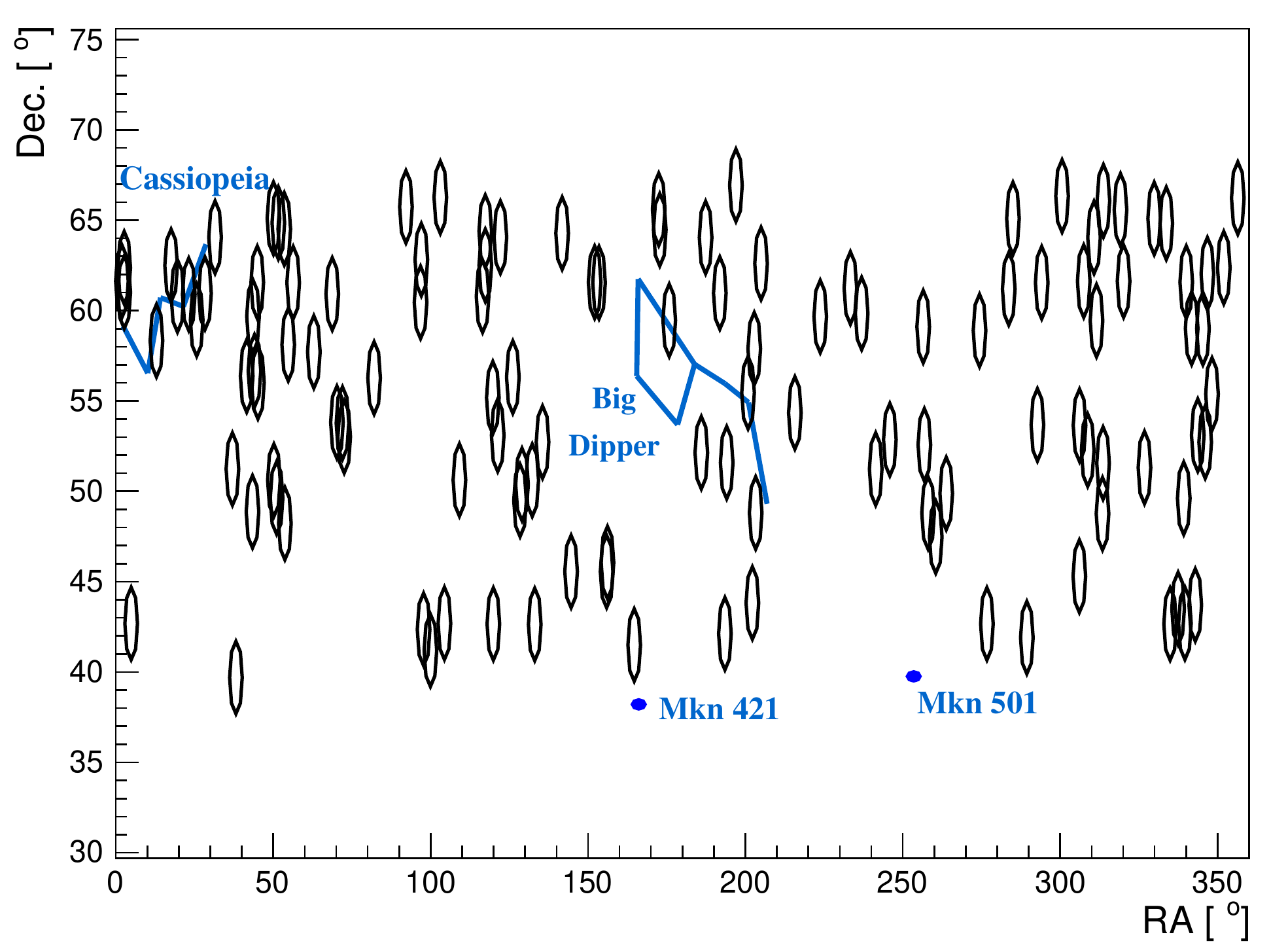}%
\caption{Celestial coordinates for the 122 track events found in the data. The size of the circles shows the average angular resolution of the tracker. The muon signals reported on this work are dominated by those of scattered atmospheric muons, the figure illustrates the field of view accessible with our analysis. The location of the Cassiopeia and Big Dipper constellations are shown as reference, as well as the Mkn 421 and Mkn 501 AGNs. \label{fig:direct}}
\end{figure}

To further motivate the capabilities of our detection method, Figure \ref{fig:direct} shows the projection of the reconstructed muon trajectories in celestial coordinates. The background of Figure \ref{fig:direct} shows as reference the Cassiopeia and Big Dipper constellations, as well as the active galaxies Mkn 421 and Mkn 501. Although our detected signals are dominated by scattered muons, the figure is used to illustrate the field of view that could be used in the future once we are able to efficiently separate the high energy signals in our data.

\section{Conclusions \label{Sec::5}}

We report on the development of a strategy to reconstruct horizontal track-like trajectories with the HAWC array of 300 water Cherenkov detectors, in an effort to detect Earth-skimming neutrino-induce charged leptons from the direction of the nearby Pico de Orizaba volcano. The analysis strategy used in this work requires that tracks propagate through at least four WCDs. This requirement sets the muon detection energy threshold of HAWC at approximately 2 GeV. 

A detailed background study was performed which indicated that scattering of low energy muons (the most probable energy of the muons that are scattered towards almost horizontal trajectories is of $\approx$ 4 GeV) into the horizontal acceptance of the detector overwhelms the signal, but that this background can be suppressed by selecting muons with energies above $\sim$100 GeV. A method to determine the energy of the detected muons is being developed. We anticipate that the current method for background rejection can be improved and among the options being explored is the use of a convolutional neural network to separate air shower from horizontal track-like topologies  \cite{NNNuHAWC}.

Using our simulations we can estimate the performance of the detector once we are able to separate the signals produced by muons with energies larger than 100 GeV, where the background from scattered muons would be significantly smaller than the neutrino-induced horizontal intensity according to the results presented on this work (see Figure \ref{fig:intensidadscatt}). Due to the increase in the trigger and algorithms efficiency with increasing muon energy the effective area for the full analysis region can reach values at least a couple of orders of magnitude larger. Combining this information with the model for the neutrino-induced muon intensity as a function of the muon energy from \cite{PhysRevD.18.2239}, we can anticipate that without any improvement in the current algorithms, we could detect at least a neutrino-induced muon every couple of years on average. Although this expected number of signals is small, we hope that this work can be used in the future to develop more efficient triggers and analysis methods that can increase significantly the number of detections with above ground experiments like HAWC.

\section{Acknowledgments}
We acknowledge the support from: the US National Science Foundation (NSF); the US Department of Energy Office of High-Energy Physics; the Laboratory Directed Research and Development (LDRD) program of Los Alamos National Laboratory; Consejo Nacional de Ciencia y Tecnolog\'ia (CONACyT), M\'exico, grants 271051, 232656, 260378, 179588, 254964, 258865, 243290, 132197, A1-S-46288, A1-S-22784, c\'atedras 873, 1563, 341, 323, Red HAWC, M\'exico; DGAPA-UNAM grants IG101320, IN111315, IN111716-3, IN111419, IA102019, IN110621; VIEP-BUAP; PIFI 2012, 2013, PROFOCIE 2014, 2015; the University of Wisconsin Alumni Research Foundation; the Institute of Geophysics, Planetary Physics, and Signatures at Los Alamos National Laboratory; Polish Science Centre grant, DEC-2017/27/B/ST9/02272; Coordinaci\'on de la Investigaci\'on Cient\'ifica de la Universidad Michoacana; Royal Society - Newton Advanced Fellowship 180385; Generalitat Valenciana, grant CIDEGENT/2018/034; Chulalongkorn University’s CUniverse (CUAASC) grant; Coordinaci\'on General Acad\'emica e Innovaci\'on (CGAI-UdeG), PRODEP-SEP UDG-CA-499. Thanks to Scott Delay, Luciano D\'iaz and Eduardo Murrieta for technical support.


\bibliography{HAWCNuv3arXiv}

\end{document}